\documentstyle[amssymb,preprint,prd,aps]{revtex}

\global\arraycolsep=2pt
\makeatletter
\def\fmslash{\@ifnextchar[{\fmsl@sh}{\fmsl@sh[0mu]}}
\def\fmsl@sh[#1]#2{  \mathchoice
    {\@fmsl@sh\displaystyle{#1}{#2}}    {\@fmsl@sh\textstyle{#1}{#2}}    
{\@fmsl@sh\scriptstyle{#1}{#2}}    {\@fmsl@sh\scriptscriptstyle{#1}{#2}}}
\def\@fmsl@sh#1#2#3{\m@th\ooalign{$\hfil#1\mkern#2/\hfil$\crcr$#1#3$}}
\makeatother

\begin{document}
\draft
\title{Orthogonality And Distinguishability: Criterion For Local Distinguishability
of Arbitrary Orthogonal States }
\author{Ping-Xing Chen$^{1,2}${\footnotesize \thanks{%
E-mail: pxchen@nudt.edu.cn}} and Cheng-Zu Li$^1$}
\address{1. Department of Applied Physics, National University of\\
Defense Technology,\\
Changsha, 410073, \\
P. R. China. \\
2. Laboratory of Quantum Communication and Quantum Computation, \\
University of Science and Technology of\\
China, Hefei, 230026, P. R. China}
\date{\today}
\maketitle

\begin{abstract}
We consider deeply the relation between the orthogonality and the
distinguishability of a set of arbitrary states (including multi-partite
states). It is shown that if a set of arbitrary states can be distinguished
by local operations and classical communication (LOCC), {\it \ }each of the
states can be written as a linear combination of product vectors such that
all product vectors of one of the states are orthogonal to the other states.
With this result we then prove a simple necessary condition for LOCC
distinguishability of a class of orthogonal states. These conclusions may be
useful in discussing the distinguishability of orthogonal quantum states
further, understanding the essence of nonlocality and discussing the
distillation of entanglement.
\end{abstract}

\pacs{PACS number(s): 89.70.+c, 03.65.ud }

\thispagestyle{empty}

\newpage \pagenumbering{arabic} 

One of the interesting features of non-locality in quantum mechanics is that
a set of orthogonal quantum states cannot be distinguished if only a single
copy of these states is provided and only local operations and classical
communication (LOCC) are allowed, in general. Taking the bipartite states as
an example, the procedure of distinguishing quantum states locally is: Alice
and Bob hold a part of a quantum system, which occupies one of $m$ possible
orthogonal states $\left| \Psi _1\right\rangle ,\left| \Psi _2\right\rangle
,...,\left| \Psi _i\right\rangle ,...,\left| \Psi _m\right\rangle $. Alice
and Bob know the precise form of these states, but don't know which of these
possible states they actually hold. To distinguish these possible states
they will perform some operations locally: Alice (or Bob) first measures her
part. Then she tells the Bob her measurement result, according to which Bob
measures his part. With the measurement results they can exclude some
possibilities of the system\cite{1}.

Many authors have considered some schemes for distinguishing locally between
a set of quantum states \cite{1,2,22,3,33,4,44,5}, both inseparable and
separable. Bennett et al showed that there are nine orthogonal product
states in a $3\otimes 3$ system which cannot be distinguished by LOCC\cite{2}%
. Walgate et al showed that any two multipartite orthogonal states can be
distinguished by LOCC\cite{1}. For two-qubit systems (or $2\otimes 2$
systems), any three of the four Bell states cannot be distinguished by LOCC 
\cite{3}. The distinguishability of quantum states has some close
connections with distillable entanglement\cite{62} and the information
transformation\cite{63}. On one hand, using the upper bound of distillable
entanglement, relative entropy entanglement\cite{52} and logarithmic
negativity\cite{53}, the authors in Ref \cite{3} proved that some states are
indistinguishable. On the other hand, using the rules on distinguishability
one may discuss the distillable entanglement\cite{61}. The LOCC\
distinguishability has link to the features of non-locality, obviously. So
the further analysis for distinguishability is meaningful.

The orthogonality acts as an important role in the distinguishability of a
set of possible states. A simple necessary condition for distinguishability
is each of the possible states is orthogonal to the other states. If the
states are locally orthogonal states \cite{44}, they can be distinguished
without classical communication (CC); if the states can be distinguished by
only projective measurements and CC, each possible state is a superpositions
of some orthogonal product vectors \cite{1,22}. A question is: for a set of
general LOCC distinguishable states, what is its orthogonality? In this
Letter, we will first show that if a set of arbitrary orthogonal states are
distinguishable by LOCC, each of the possible states has at least a
product-vectors-decomposition such that the product vectors of each of the
possible states are orthogonal to the other possible states. With this
result we then prove a simple necessary condition for LOCC
distinguishability of a class of orthogonal states. These conclusions may be
useful in discussing the distinguishability of orthogonal quantum states
further, understanding the essence of nonlocality\cite{z} and discussing the
distillation of entanglement.

We first take the bipartite states as an example for simplicity.. Consider $m
$ possible orthogonal states shared between Alice and Bob. Any protocol to
distinguish the $m$ possible orthogonal states can be conceived as
successive rounds of measurements and communication by Alice and Bob. Let us
suppose Alice is the first person to perform a measurement (Alice goes first 
\cite{22}), and the first round measurement by Alice can be represented by
operators $\left\{ A_{1_j}\right\} $, where $A_{1_j}^{+}A_{1_j}$ is known as
a POVM element realized by Alice \cite{p,h}, and $\sum_jA_{1_j}^{+}A_{1_j}=I.
$ If the outcome $1_j$ occurs, then the given $\left| \Psi \right\rangle $
becomes $A_{1_j}\left| \Psi \right\rangle ,$ up to normalization. After
communicating the result of Alice's measurement to Bob, he carries out a
measurement and obtain outcome $1_k$. The given possible state $\left| \Psi
\right\rangle $ becomes $A_{1_j}\otimes B_{1_k}(1_j)\left| \Psi
\right\rangle $, where $B_{1_k}(1_j)$ is an arbitrary measurement operator
of Bob which depend on the outcome $1_j$ of Alice's measurement. After N
rounds of measurements and communication, there are many possible outcomes
which correspond to many measurement operators acting on the Alice and Bob's
Hilbert space. Each of these operators is a product of the N sequential and
relative operators, $A_{N_j}(1_j,1_k,...,(N-1)_k)\otimes
B_{N_k}(1_j,1_k,...,(N-1)_k,N_j)...A_{2_j}(1_j,1_k)\otimes
B_{2_k}(1_j,1_k,2_j)A_{1_j}\otimes B_{1_k}(1_j),$ carried out by Alice and
Bob. We denote these operators as $\left\{ A_i\otimes B_i\right\} ,$ where, $%
A_i\otimes $ $B_i$ denotes one of these operators, which represent the
effects of the N measurements and communication. If the outcome $i$ occurs,
the given $\left| \Psi \right\rangle $ becomes:

\begin{equation}
A_i\otimes B_i\left| \Psi \right\rangle
\end{equation}
The probability $p_i$ Alice and Bob gain outcome $i$ is

\begin{equation}
p_i=\left\langle \Psi \right| A_i^{+}\otimes B_i^{+}A_i\otimes B_i\left|
\Psi \right\rangle ,
\end{equation}
and

\begin{equation}
\sum_iA_i^{+}\otimes B_i^{+}A_i\otimes B_i=I.
\end{equation}
Suppose we define:

\begin{equation}
E_i=A_i^{+}\otimes B_i^{+}A_i\otimes B_i,  \label{55}
\end{equation}
then $E_i$ is a positive operator and that $\sum_iE_i=I.$ $E_i$ is same as
the known POVM element. In fact, $A_i$ can be written in the form \cite{p}

\begin{equation}
A_i=U_{A2}f_{Ai}U_{A1},  \label{s}
\end{equation}
or

\begin{eqnarray}
A_i &=&c_1^i\left| \varphi _1^{\prime i}\right\rangle \left\langle \varphi
_1^i\right| +\cdots +c_{m_a^i}^i\left| \varphi _{m_a^i}^{\prime
i}\right\rangle \left\langle \varphi _{m_a^i}^i\right| ;  \label{7} \\
0 &\leq &c_j^i\leq 1,j=1,\cdots ,m_a^i.  \nonumber
\end{eqnarray}
$\ $Where $f_{Ai}$ is a diagonal positive operator and a filtration \cite{p}
which changes the relative weights of components $\left| \varphi
_1^i\right\rangle ,\cdots ,\left| \varphi _{n_i}^i\right\rangle $; $%
U_{A2},U_{A1}$ are two unitary operators; $\left\{ \left| \varphi _1^{\prime
i}\right\rangle ,\cdots ,\left| \varphi _{m_a^i}^{\prime i}\right\rangle
\right\} $ and $\left\{ \left| \varphi _1^i\right\rangle ,\cdots ,\left|
\varphi _{m_a^i}^i\right\rangle \right\} $ are two set of orthogonal Alice's
vectors, and similarly for $B_i.$

\begin{eqnarray}
B_i &=&d_1^i\left| \eta _1^{\prime i}\right\rangle \left\langle \eta
_1^i\right| +\cdots +d_{m_b^i}^i\left| \eta _{m_b^i}^{\prime i}\right\rangle
\left\langle \eta _{m_b^i}^i\right|  \label{77} \\
0 &\leq &d_j^i\leq 1,j=1,\cdots ,m_b^i.  \nonumber
\end{eqnarray}
where \{$\left| \eta _1^{\prime i}\right\rangle ,\cdots ,\left| \eta
_{m_b^i}^{\prime i}\right\rangle \}$ and $\left\{ \left| \eta
_1^i\right\rangle ,\cdots ,\left| \eta _{m_b^i}^i\right\rangle \right\} $
are two set of orthogonal Bob's vectors.

From Eq.(\ref{55}), Eq.(\ref{7}) and Eq.(\ref{77}), we can represent $E_i$
in the form

\begin{eqnarray}
E_i &=&(a_1^i\left| \varphi _1^i\right\rangle _A\left\langle \varphi
_1^i\right| +\cdots +a_{m_a^i}^i\left| \varphi _{m_a^i}^i\right\rangle
_A\left\langle \varphi _{m_a^i}^i\right| )\otimes  \label{e} \\
&&(b_1^i\left| \eta _1^i\right\rangle _B\left\langle \eta _1^i\right|
+\cdots +b_{m_b^i}^i\left| \eta _{m_b^i}^i\right\rangle _B\left\langle \eta
_{m_b^i}^i\right| )  \nonumber \\
0 &\leqslant &a_{m_a^i}^i\leqslant 1,0\leqslant b_{m_b^i}^i\leqslant
1;1\leqslant m_a^i\leqslant N_a,1\leqslant m_b^i\leqslant N_b
\end{eqnarray}
where $N_a,N_b$ is the dimensions of Alice's and Bob's Hilbert space,
respectively.

The discussion above means that: whatever Alice and Bob choose to do by
LOCC, their final actions will be described by a set of positive operators $%
\left\{ E_i\right\} .$ This result is useful to the following discussions.

Theorem 1. If a set of $m$ orthogonal states $\left\{ \left| \Psi
_i\right\rangle \right\} $ is perfectly distinguishable by LOCC, there is
surely a set of {\it product vectors }(PV) such that each state $\left| \Psi
_i\right\rangle $ is a superposition of some of these {\it product vectors}
as follows:

\begin{equation}
\left| \Psi _i\right\rangle =\left| \Phi _i^1\right\rangle _A\left| \xi
_i^1\right\rangle _B+\cdots +\left| \Phi _i^{m^i}\right\rangle _A\left| \xi
_i^{m^i}\right\rangle _B;  \label{th}
\end{equation}
and each product vector $\left| \Phi _i^{k^i}\right\rangle _A\left| \xi
^{k^i}\right\rangle _B$ ($1\leqslant k^i\leqslant m^i$) belongs to only a
state $\left| \Psi _i\right\rangle ,i.e.,$

\begin{equation}
\left\langle \Phi _i^{k^i}\right| \left\langle \xi _i^{k^i}\right| \Psi
_j\rangle =0,\text{ for all }i\neq j;  \label{h}
\end{equation}
\begin{equation}
\left\langle \Phi _i^{k^i}\right| \left\langle \xi _i^{k^i}\right| \Psi
_i\rangle \neq 0,  \label{t}
\end{equation}
where $m^i$ is a positive integral number.

Proof: If a set of states is reliably distinguishable by LOCC, there must be
a complete set of POVM element $\left\{ E_i\right\} $ representing the
effect of all measurements and communication, such that if every outcome $i$
occurs Alice and Bob know with certainty that they were given the state $%
\left| \Psi _i\right\rangle $. This means that:

\begin{equation}
\left\langle \Psi _i\right| E_{i(s)}\left| \Psi _i\right\rangle \neq 0;
\label{8}
\end{equation}
\begin{equation}
\left\langle \Psi _j\right| E_{i(s)}\left| \Psi _j\right\rangle =0,j\neq i.
\label{88}
\end{equation}
In a simple way, we can say that a element $E_i$ can ``indicate'' $\left|
\Psi _i\right\rangle $ and only $\left| \Psi _i\right\rangle .$ Of course a
possible state may be ``indicated'' by more than one POVM element $E_i.$ $%
E_{i(s)}$ denotes all $E_i$ ``indicating'' $\left| \Psi _i\right\rangle $.
Note that because the non-projective measures and the classical
communication between Alice and Bob are allowed, some POVM elements in $%
\left\{ E_i\right\} $ can be not orthogonal to others.

From the general expression of a operator $A_i$ in Eq.(\ref{7}), it follows
that a operator $A_i$ in a POVM element $E_i$ in Eq.(\ref{55}) can be
carried out by the following operators: 1). do projective operation $P_A^i,$

\begin{equation}
P_A^i=\left| \varphi _1^i\right\rangle \left\langle \varphi _1^i\right|
+\cdots +\left| \varphi _{n_i}^i\right\rangle \left\langle \varphi
_{n_i}^i\right| ,  \label{9}
\end{equation}
which projects out the Alice's component $\left| \varphi _1^i\right\rangle
,\cdots ,\left| \varphi _{n_i}^i\right\rangle $ in a possible state $\left|
\Psi _i\right\rangle $ (if $\left| \Psi \right\rangle $=$\left|
0\right\rangle _A\left| 0\right\rangle _B$+$\left| 1\right\rangle _A\left|
1\right\rangle _B$,we say $\left| \Psi \right\rangle $ have components $%
\left| 0\right\rangle _A\left| 0\right\rangle _B$ and $\left| 1\right\rangle
_A\left| 1\right\rangle _B;\left| \Psi \right\rangle $ have Alice's
components $\left| 0\right\rangle _A$ and $\left| 1\right\rangle _A$)$;$ 2).
do local filter operation \cite{pp} which changes the relative weights of
the component $\left| \varphi _1^i\right\rangle ,\cdots ,\left| \varphi
_{n_i}^i\right\rangle $ in a possible state $\left| \Psi _i\right\rangle $;
3). do a local unitary operation which transfers the Alice's bases from $%
\left\{ \left| \varphi _1^i\right\rangle ,\cdots ,\left| \varphi
_{n_i}^i\right\rangle \right\} $ to $\left\{ \left| \varphi _1^{\prime
i}\right\rangle ,\cdots ,\left| \varphi _{n_i}^{\prime i}\right\rangle
\right\} ,$ and similarly for $B_i.$ So if $E_i$ ``indicates'' a state $%
\left| \Psi _i\right\rangle ,$ i.e., Eq. (\ref{8}) holds, the state $\left|
\Psi _i\right\rangle $ should have all or part of the following components: 
\begin{equation}
\left| \varphi _1^i\right\rangle _A\left| \eta _1^i\right\rangle _B,\cdots
,\left| \varphi _1^i\right\rangle _A\left| \eta _{m_b^i}^i\right\rangle
_B,\cdots ,\left| \varphi _{m_a^i}^i\right\rangle _A\left| \eta
_1^i\right\rangle _B,\cdots ,\left| \varphi _{m_a^i}^i\right\rangle _A\left|
\eta _{m_b^i}^i\right\rangle _B.  \label{10}
\end{equation}
If $E_i$ ``indicates'' only the state $\left| \Psi _i\right\rangle ,$ i.e.,
Eq. (\ref{88}) holds, each product vector in (\ref{10}) should be orthogonal
to the other states $\left| \Psi _j\right\rangle ,$ for all $j\neq i.$ We
may say that $E_i$ also ``indicates'' each product vector in (\ref{10})
which belongs to only the state $\left| \Psi _i\right\rangle .$

Because of the completeness of $\left\{ E_i\right\} $, which assures that
each product vector in all possible states can be indicated by a POVM
element, and the necessity of reliably distinguishing the possible states,
which asks a POVM element ``indicates'' the product vectors of only a
possible state, each state of the $m$ possible states must be a
superposition of many product vectors each of which is orthogonal to the
other possible states. This ends the proof.

The above theorem 1 shows that if a set of possible states are LOCC
distinguishable, not only that these possible states should be orthogonal,
but also each possible state can be written as a linear combination of
product vectors such that each product vector of a possible state $\left|
\Psi _i\right\rangle $ should be orthogonal to the other possible states.
There are two ``opposite'' cases \cite{22,33}: 1. entanglement may increase
the local indistinguishability of orthogonal states. An example is: $nm$
orthogonal states of a $n\otimes m$ system cannot be perfectly LOCC
distinguishable if at least one of the states is entangled (see \cite{33});
2. entanglement may increase the local distinguishability of orthogonal
states. An example is: the set {\it S }containing states (without
normalization):

\begin{eqnarray}
\left| \Psi _1\right\rangle  &=&\left| 00\right\rangle +w\left|
11\right\rangle +w^2\left| 22\right\rangle ;\left| \Psi _2\right\rangle
=\left| 00\right\rangle +w^2\left| 11\right\rangle +w\left| 22\right\rangle ;
\label{ss} \\
\left| \Psi _3\right\rangle  &=&\left| 01\right\rangle +\left|
12\right\rangle +\left| 20\right\rangle ,  \nonumber
\end{eqnarray}
is LOCC\ distinguishable ($w$ is a unreal cube root of unity). But the
states, $\left| \Psi _1\right\rangle ,\left| \Psi _2\right\rangle $ and $%
\left| \Psi _3^{\prime }\right\rangle =\left| 01\right\rangle $ are not LOCC
distinguishable (see \cite{33}). In fact, entanglement as a potential
non-local ``resource'' may increase the distinguishability of the states (to
distinguish a set of states, the entanglement of these states may be lost,
in general. This means entanglement can be used to distinguish states as to
do teleportation et al). But on other hand, a entangled state contains more
product vectors. So the entangled state increases the requirement for
orthogonality as shown in the above theorem 1 and then may increase the
indistinguishability of the states.

Employing theorem 1 we can discuss the LOCC distinguishability of orthogonal
states further. First, the above discussions and theorem 1 are fit to the
multipartite systems obviously. The generalisation of the theorem 1 can be
expressed as:

{\it if a set of multi-partite possible states are LOCC distinguishable,
each possible state can be written as a linear combination of product
vectors such that each product vector of a possible state is orthogonal to
the other possible states}. 

Then we will follow a especially simple criterion for distinguishability of
a class of orthogonal states. To achieve this, we define a concept of {\it %
Schmidt number}. If a pure state $\left| \Psi \right\rangle $ have following
Schmidt decomposition:

\begin{equation}
\left| \Psi \right\rangle =\sum_{i=1}^l\sqrt{p_i}\left| \phi _i\right\rangle
_A\left| \eta _i\right\rangle _B,\qquad p_i>0,\qquad \sum_{i=1}^lp_i=1
\label{2}
\end{equation}
where $\left| \phi _i\right\rangle _As$ and $\left| \eta _i\right\rangle _Bs$
are orthogonal bases of Alice and Bob, respectively, we say $\left| \Psi
\right\rangle $ has {\it Schmidt number} $l.$

Theorem 2: Let $\left\{ \left| \Psi _i^{AB}\right\rangle \right\} $ is $%
mn-m^{\prime }$ orthogonal states of an $m\otimes n$ system, if at least one
of the states has Schmidt numbers bigger than $m^{\prime }+1,$ the states $%
\left| \Psi _i^{AB}\right\rangle s$ are not perfectly distinguishable by
LOCC.

Proof : Each state of the $\left\{ \left| \Psi _i^{AB}\right\rangle \right\} 
$ should include linearly independent product vectors(LIPV) not less than
its Schmidt numbers$.$ If at least one state (note it as $\left| \Psi
_{i^{\prime }}^{AB}\right\rangle )$ among the $\left| \Psi
_i^{AB}\right\rangle s$ has Schmidt numbers bigger than $m^{\prime }+1,$ an
assumption of local distinguishability of $\left| \Psi _i^{AB}\right\rangle s
$ implies that the state $\left| \Psi _{i^{\prime }}^{AB}\right\rangle $ is
a superposition of more than $m^{\prime }+1$ LIPVs each of which is
orthogonal to the other states $\left| \Psi _i^{AB}\right\rangle s(i\neq
i^{\prime }).$ Thus the states $\left| \Psi _i^{AB}\right\rangle s(i\neq
i^{\prime })$ and the LIPVs of $\left| \Psi _{i^{\prime }}^{AB}\right\rangle 
$ form a set of linearly independent vectors of an $m\otimes n$ system. The
number of these linearly independent vectors is bigger than $mn-m^{\prime
}-1+m^{\prime }+1=mn.$ This is impossible for an $m\otimes n$ system. So the
states $\left| \Psi _i^{AB}\right\rangle s$ are not perfectly
distinguishable by LOCC. This ends the proof.

Since we cannot define the Schmidt numbers of a multi-partite pure state, in
general, the theorem 2 above cannot be generalized to multi-partite states
directly. However, any pure state has a product-vectors-decomposition with
the least number of product vectors, if we replace ''Schmidt numbers'' in
the theorem 2 by ''the least numbers of product vectors'', the theorem 2 can
be generalized into multi-partite states. For example, an 3-qubits system
owns 7 possible orthogonal states. If one of the possible states is a $%
\left| W\right\rangle $ state

\begin{equation}
\left| W\right\rangle =\frac 1{\sqrt{3}}(\left| 100\right\rangle +\left|
010\right\rangle +\left| 011\right\rangle ),  \label{w}
\end{equation}
these possible are not perfectly distinguishable by LOCC since a $\left|
W\right\rangle $ state has at least three LIPVs \cite{w}.

The theorem 2 is powerful to check LOCC indistinguishability of the
orthogonal states the number of which is equal to or near to the dimensions
of the quantum system. For example, one can get by theorem 2 easily that if
a full orthogonal basis can be LOCC distinguished all vectors must be
product, as shown in Ref. \cite{33}.

To conclude, we have considered deeply the relation between the
orthogonality and the distinguishability, and shown that if a set of
possible multi-partite orthogonal states are LOCC distinguishable, each of
the possible states has at least a product-vectors-decomposition such that
the product vectors of each of the possible states are orthogonal to the
other possible states. Based on our result one can discuss the
distinguishability of orthogonal states further. We also present a simple
necessary condition for distinguishability of a class of orthogonal quantum
states. These results come directly from the limits on local operations, not
from the upper bound of distillable entanglement\cite{3}, So we believe that
they may be useful in understanding the essence of nonlocality. On the other
hand, the distillation of entanglement and local distinguishability are
closely related as shown in Ref.\cite{3,33,61}, so our results may be
helpful for calculating the distillable entanglement or the bound of
distillable entanglement. The further works may be the applications of these
results.

\acknowledgments    We would like to thank J. Finkelstein, A. Sen(De) and U.
Sen for their helpful suggestions by E-mail and Guangcan Guo for his help to
this work.

\end{document}